\documentclass[namedreferences]{SolarPhysics}
\usepackage[optionalrh]{spr-sola-addons} % For Solar Physics
\usepackage{graphicx}        % For eps figures, newer & more powerfull
\usepackage{color}           % For color text: \color command
\usepackage{url}             % For breaking URLs easily trough lines
\renewcommand{\vec}[1]{ {\mathbf #1} }

\newcommand{\dv}{~{\mathrm d}^3 x}

\newcommand{\avec}{ \vec A}

\newcommand{\bb}{ \vec B}

\newcommand{\adv}{    {\it Adv. Space Res.}}

\newcommand{\aap}{    {\it Astron. Astrophys.}}

\newcommand{\aj}{     {\it Astronom. J.}}
\newcommand{\apj}{    {\it Astrophys. J.}}

\newcommand{\grl}{    {\it Geophys. Res. Lett.}}

\newcommand{\jgr}{    {\it J. Geophys. Res.}}

\newcommand{\pasj}{   {\it Pub. Astron. Soc. Japan}}

\newcommand{\solphys}{{\it Solar Phys.}}

\newcommand{\ajsr}{   {\it Astrophys. J. Suppl. Ser.}}

\begin{document}

\begin{article}

\begin{opening}

\title{Evidence of Magnetic Helicity in Emerging Flux and Associated Flare}

\author{R.~\surname{Chandra}\sep
        B.~\surname{Schmieder}\sep
        G.~\surname{Aulanier}\sep
        J.M.~\surname{Malherbe}\
       }
\runningauthor{R. Chandra et al.} \runningtitle{Evidence of magnetic
helicity}

\institute{R. Chandra \sep B. Schmieder \sep  G. Aulanier \sep J.M. Malherbe\\
Observatoire de Paris, LESIA, UMR8109 (CNRS), F-92195 Meudon Principal Cedex, France\\
\medskip
R. Chandra\\
email: \url{chandra.ramesh@obspm.fr}\\
\medskip
B. Schmieder\\
email: \url{brigitte.schmieder@obspm.fr}\\
\medskip
G. Aulanier\\
email: \url{guillaume.aulanier@obspm.fr}\\
\medskip
J.M. Malherbe\\
email: \url{jean-marie.malherbe@obspm.fr}\\
\medskip
R. Chandra \\
Uttarakhand Department of Science and Technology C/O Office of the District Magistrate, Nainital, India  \\
             }

%%%%%%%%%%%%%%%%%%%%%%%%%%%%%%%%%%%%%%%%%%%%%%%%%%%%%%%%%%%%%%%%%%%%%%%%%%%%%%%%%%%%%%

\begin{abstract}
The aim of this paper is to  look at the magnetic helicity structure
of an emerging active region and show that both emergence and flaring
signatures are consistent with a same sign for magnetic helicity.
We present a multi-wavelength analysis of an M1.6 flare occurring in
the active region NOAA 10365 on 27 May, 2003, in which a large new bipole emerges in
a decaying active region. The diverging flow
pattern and the ``tongue'' shape of the magnetic field  in the
photosphere with elongated polarities are highly suggestive of the
emergence of a twisted flux tube. The orientation of these tongues indicates the
emergence of a flux tube with a right hand twist, i.e. positive magnetic helicity.
The flare signatures in the chromosphere are ribbons observed
in H$\alpha$ by the MSDP spectrograph in the Meudon solar tower and
in 1600 \AA\ by TRACE. These ribbons have a `J' shape and are
shifted along the inversion line. The pattern of these ribbons
suggests that the flare was triggered by magnetic reconnection at coronal heights
below a twisted flux tube of positive helicity, corresponding to that of the
observed emergence. It is the first time that such a consistency
between  the signatures of  the emerging flux through the
photosphere and flare ribbons is clearly identified in observations.
Another type of ribbons observed during the flare at the periphery of the active region by
the MSDP and SOHO/EIT are related to the existence of a null point, which is found
high in the corona in a potential field extrapolation. We discuss
the interpretation of these secondary brightenings in terms of the ``breakout'' model and
in terms of plasma compression/heating within large-scale separatrices.

\end{abstract}
%\keywords{Flares, Helicity, Magnetic fields, Corona}
\end{opening}
%-------------------------------------------------

%%%%%%%%%%%%%%%%%%%%%%%%%%%%%%%%%%%%%%%%%%%%%%%%%%%%%%%%%%%%%%%%%%%%%%%%%%%%%%%%%%%%%%%%%%%%%%%%%%%%%%%

\section{Introduction}
     \label{S-Introduction}

It is believed that flare energy originates from the free energy
stored in  non-potential magnetic structures. Magnetic reconnection
is considered to play an important role for the sudden release of
energy in  solar flares. Several conditions such as: flux emergence,
high magnetic field gradient, flux cancellation etc. are commonly
found near flare locations.

Magnetic complexity plays an important role for strong solar
activity, as studied by many authors (i.e. \opencite{Schmieder05} ;
\opencite{Sammis00}; \opencite{Antiochos99}). The magnetic
complexity is commonly due to slow or fast evolution of the magnetic
configuration. Shear and flux emergence are responsible of such changes 
in the magnetic field. \inlinecite{Hagyard90} have shown that the magnetic
field in flaring locations is strongly sheared. Large scale
shear in the magnetic field can  be built up through the slow motion
of footpoints stretching the length of  loops. This evolution can
progress through stable field configurations \cite{Roudier08}.
However, magnetic shear alone is not a sufficient condition.
Flux emergence has long been considered as an important condition to
trigger solar flares (\opencite{Heyvaerts77}; \opencite{Martin82};
\opencite{Shibata92}; \opencite{Schmieder94};
\opencite{Choudhary98}; \opencite{Chen00}; \opencite{Zhangy08} and
references therein). A new magnetic flux tube emerges from
below the photosphere, interacts with the existing flux tubes, and
the reconnection occurs in the current sheet, which forms between
the old and new fluxes \cite{Heyvaerts77}. The magnetic flux may emerge 
already twisted \cite{Leka96}. 
Sub-photospheric motions are the drivers of free energy storage
and the emerging flux provides the trigger mechanism for impulsive
energy release (\opencite{Zhao04}; \opencite{Mason06}).

Magnetic helicity is a parameter indicating the degree of twist
and writhe in an active region (see the reviews by \opencite{Demoulin07};
\opencite{Demoulin08}). The helicity pattern of the Sun
has characteristic features governed by an hemispheric rule:
positive/negative magnetic helicity is common in the South/North
hemisphere. \inlinecite{Pevtsov02} summarized the different possible
signatures of magnetic helicity observed on the Sun: sunspot whorls,
filament chirality, sigmoids, flare ribbons.
Appearance of magnetic ``tongues'' in an emerging active region as first discussed
by \inlinecite{Lopez00} is an important signature, which tells about the helicity sign
of the active region. We will use these proxies to derive
the magnetic helicity of our active region.
Sometimes there are mixed magnetic helicity zones in the active region.
Therefore it is not always necessary that the helicity sign be the same
in all the observational features (\opencite{Green07}; \opencite{Chandra09}).

The analysis of magnetic topology of an active region gives us
information on possible locations of magnetic
reconnection: null point (locations where magnetic field vanishes), separatrices (which
separate different connectivity domain), Quasi-separatrices layers (QSL) (the region
where there is a drastic change in field line linkage). QSL is a generalization of
this concept proposed by \inlinecite{Priest95} and applied to the modeling of `J' shaped 
two-ribbon flares by \inlinecite{Demoulin96}.

In this paper, we study the evolution and magnetic configuration
of the M1.6 flare that occurred in the NOAA 10365 active region  on
27 May, 2003, using multiwavelength observations.
We present the instrumentation and observations in
Section 2. Section 3 describes the evolution of the magnetic field in
the active region. The description of the flare at different
wavelengths is presented in Section 4. In Section 5 we explain the relationship
between the magnetic helicity of the twisted emerging flux tube and the magnetic reconnection
leading to the M1.6 flare with `J' shape main ribbons. 
In Section 6, we show the results of a linear force-free field (lfff) model of the large scale active region configuration and provide an interpretation to the presence of semicircular brightenings (secondary ribbons) in its periphery.   
We finally conclude that it is the first time that signatures of
magnetic helicity in the photosphere (presence of tongues in longitudinal magnetic field maps),
 in the chromosphere (`J' shaped ribbons) and in the corona (hard
 X-ray sources) are consistent with the existence of an emerging
twisted magnetic flux tube and reconnection in the corona.

%%%%%%%%%%%%%%%%%%%%%%%%%%%%%%%%%%%%%%%%%%%%%%%%%%%%%%%%%%%%%%%%%%%%%%%%%%%%%%%%%%%%%%%

\section{Instrumentation and Observations} %%%%%%%%%%%%%%%%%%%%%%%%%%%%%%%%%%%%%%%%
      \label{S-general}

The M1.6 flare occurring in the active region NOAA 10365 located S08 W10 on 
27 May, 2003 was the target of an international campaign. This flare is associated with
a coronal mass ejection (CME) observed by {\it Large Angle Spectroscopic Coronagraph} (LASCO, \opencite{Brueckner95}) 
at 06:50 UT. The complex morphology
of the ribbons observed at the Meudon solar tower was intriguing and initiated
this study.

%%%%%%%%%%%%%%%%%%%%%%%%%%%%%%%%%%%%%%%%%%%%%%%%%%%%%%%%%%%%%%%%%%%%%%%%%%%%%%%%%%%%%%%%%%%%%%%%%%%%%%%%
\begin{figure} %%%%%%%%%%%%%%%%%% FIGURE 1

\centerline{\hspace*{0.015\textwidth}
               \includegraphics[width=0.80\textwidth,clip=]{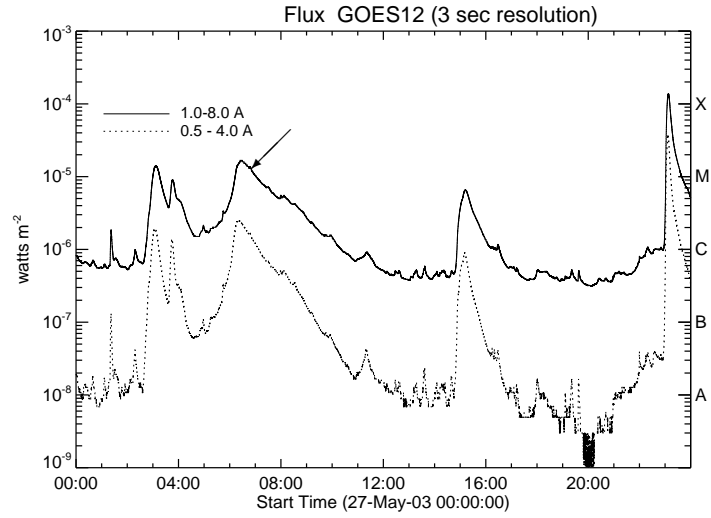}
              $\color{black} \put(-165,150){\vector(-1,-1){15}}$
        }

     \vspace{-0.02\textwidth}    % Shift back to the panel bottom

\caption{Temporal Evolution of the flare in GOES 12 X-ray (0.5-4) \AA\ and
(1-8) \AA\ on May 27, 2003. The studied M1.6 flare is shown by an arrow.}
   \label{F-4panels}
   \end{figure}

Figure \ref{F-4panels} shows the temporal evolution of the flare
observed by GOES 12. The eruption onset starts at 05:40 UT with a
flux of 3.5$\times10^{-6}$ Wm$^{-2}$. The maximum of flux is reached
around 06:15 UT with a value in X-rays (1-8 \AA\ ) of 1.6$\times
10^{-5}$ W m$^{-2}$. This is a long duration event that ended around
10:00 UT.

%%%%%%%%%%%%%%%%%%%%%%%%%%%%%%%%%%%%%%%%%%%%%%%%%%%%%%%%%%%%%%%%%%%%%%%%%%%%%%%%%%%%%%%%%%%%%%%%%%%%%

\begin{figure} %%%%%%%%%%%%%%%%%% FIGURE 2

\centerline{\hspace*{0.015\textwidth}
               \includegraphics[width=0.76\textwidth,clip=]{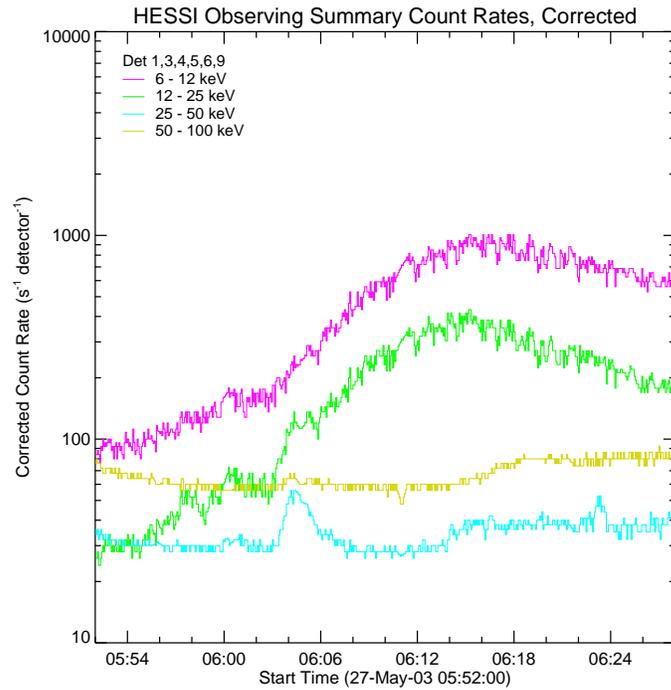}
              }

     \vspace{-0.02\textwidth}    % Shift back to the panel bottom

\caption{Temporal Evolution of the flare in HXR observed by RHESSI 
at 6-12, 12-15, 25-50 and 50-100 keV, respectively.}
   \label{rhessi}
   \end{figure}
%%%%%%%%%%%%%%%%%%%%%%%%%%%%%%%%%%%%%%%%%%%%%%%%%%%%%%%%%%%%%%%%%%%%%%%%%%%%%%%%%%%%%%%%%%%%%%%%

Figure \ref{rhessi} presents the temporal evolution of the flare in
hard X-rays (HXR) observed by  {\it Reuven Ramaty High-Energy Solar Spectroscopic Imager} (RHESSI). The 
time profiles show that the X-ray
emissions above 25 keV are very low. The time profiles indicate that
the thermal component is dominant in the flare. 
The two low energy band emissions slightly increase
after 05:40 UT until 06:15 UT, with an impulsive event at 06:04 UT
detectable in the higher energy band (25-50 keV).

In this study, we use data from {\it Solar and Heliospheric Observatory} (SOHO)/ {\it Michelson Doppler Imager} (MDI) (time cadence of 96 min and pixel size
of 1.98$''$; \opencite{Scherrer95} ) and EIT (time cadence of 12 min to 6 hour and pixel size of
2.5$''$; \opencite{Delaboudiniere95}), {\it Transition Region and Coronal Explorer} (TRACE) 1600 \AA\ (time 
cadence of 6 min and pixel size of 0.5$''$,
\opencite{Handy99}); RHESSI \cite{Lin02} as well as Multi channel Subtractive Double
Pass spectrograph (MSDP) of the Meudon solar tower.

The M1.6 flare  has been observed in the
H$\alpha$ line (6563 \AA) with the MSDP.
The entrance slit of the spectrograph covers an elementary field-of-view
of 72$''\times465''$  with a pixel size of 0.5$''$ . The final
field-of-view of the images is 295 $''\times460''$. The exposure time is 40 ms. We performed
consecutive sequences of 5 images with a cadence of 1 min.
Using the MSDP technique (\opencite{Mein77}; \opencite{Mein91})
the field-of-view is recorded simultaneously in nine different
wavelength intervals around H$\alpha$, line center, separated by 0.3 \AA. 
Interpolating with spline functions the observed intensity in these
images, we are able to construct H$\alpha$ profiles in each point of
the observed field-of-view. A mean or reference disk profile is
obtained by averaging over a quiet
region on the disk (flat field sequence).

%%%%%%%%%%%%%%%%%%%%%%%%%%%%%%%%%%%%%%%%%%%%%%%%%%%%%%%%%%%%%%%%%%%%%%%%%%%%%%%%%
\begin{figure} %%%%%% FIGURE 3
\vspace*{-0.8cm}
\centerline{\hspace*{-0.02\textwidth}
               \includegraphics[width=0.95\textwidth,angle=-90,clip=]{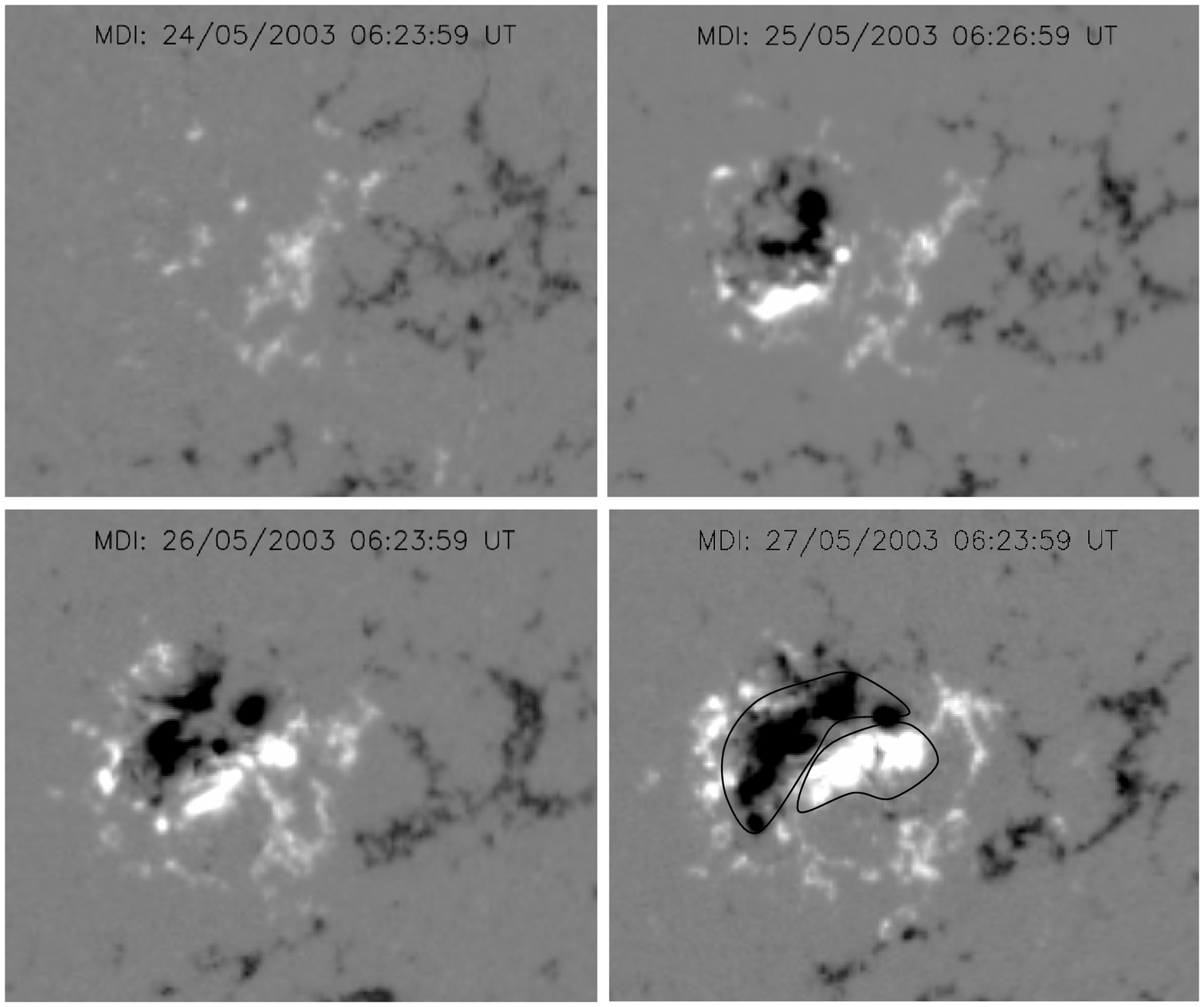}
$\color{black} \put(-385,-55){\bf\large (a)}$         
$\color{black} \put(-228,-55){\bf\large (b)}$
$\color{black} \put(-392,-192){\bf\large (c)}$
$\color{black} \put(-235,-192){\bf\large (d)}$
\vspace*{-0.5cm}
                }
\caption{ Evolution of the magnetic field of active region NOAA
10365. Magnetograms from 24 May, 2003 to 27 May, 2003 show
the emergence of high intensity flux.
The contour over the magnetogram of 27 May, 2003 shows the two 
magnetic tongues. The field of view of
the images is 300$''\times 250''$ . North is up and West is to
the right.}
\label{mdi}
\end{figure}

%%%%%%%%%%%%%%%%%%%%%%%%%%%%%%%%%%%%%%%%%%%%%%%%%%%%%%%%%%%%%%%%%%%%%%%%%%%%%%%%%%%%%%%%%%%%%%%%%%%%%%

\section{Evolution of the Photospheric Magnetic Configuration} %%%%%%%%%%%%%%%%
      \label{S-general}

\subsection{Global Pattern of the emerging flux}

In order to understand the cause of the flare, we need to study the
magnetic configuration of the active region.
Using MDI magnetograms we were able to follow its evolution 
during several days in May, 2003 (Figure
\ref{mdi}). On May 24, this region 
had a weak field bipolar configuration
with a leading negative polarity and
and a diffuse following positive polarity.
This corresponds to a decaying active region.

%%%%%%%%%%%%%%%%%%%%%%%%%%%%%%%%%%%%%%%%%%%%%%%%%%%%%%%%%%%%%%%%%%%%%%%%%%%%%%%%%%%%%%%%%%

\begin{figure}    %%%%%%%%%%%%%%%%%% FIGURE 4
                                % includes the two top panels
%\vspace{-0.05\textwidth}    % Shift back to the panel bottom
   \centerline{\hspace*{0.015\textwidth}
               \includegraphics[width=0.45\textwidth,clip=]{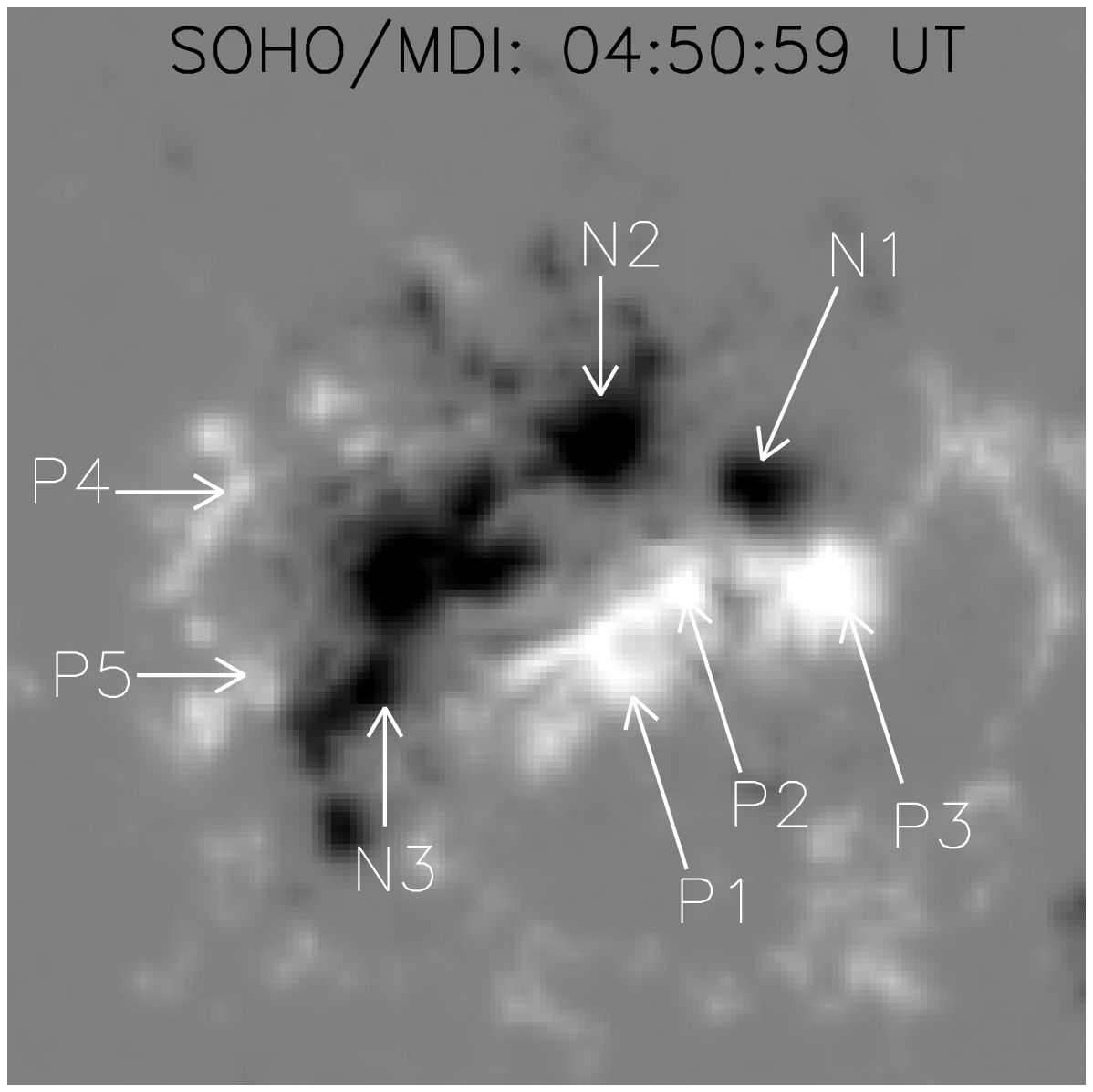}
$\color{black} \put(-160,145){\bf\large (a)}$         
               %\hspace*{-0.47\textwidth}
               \includegraphics[width=0.45\textwidth,clip=]{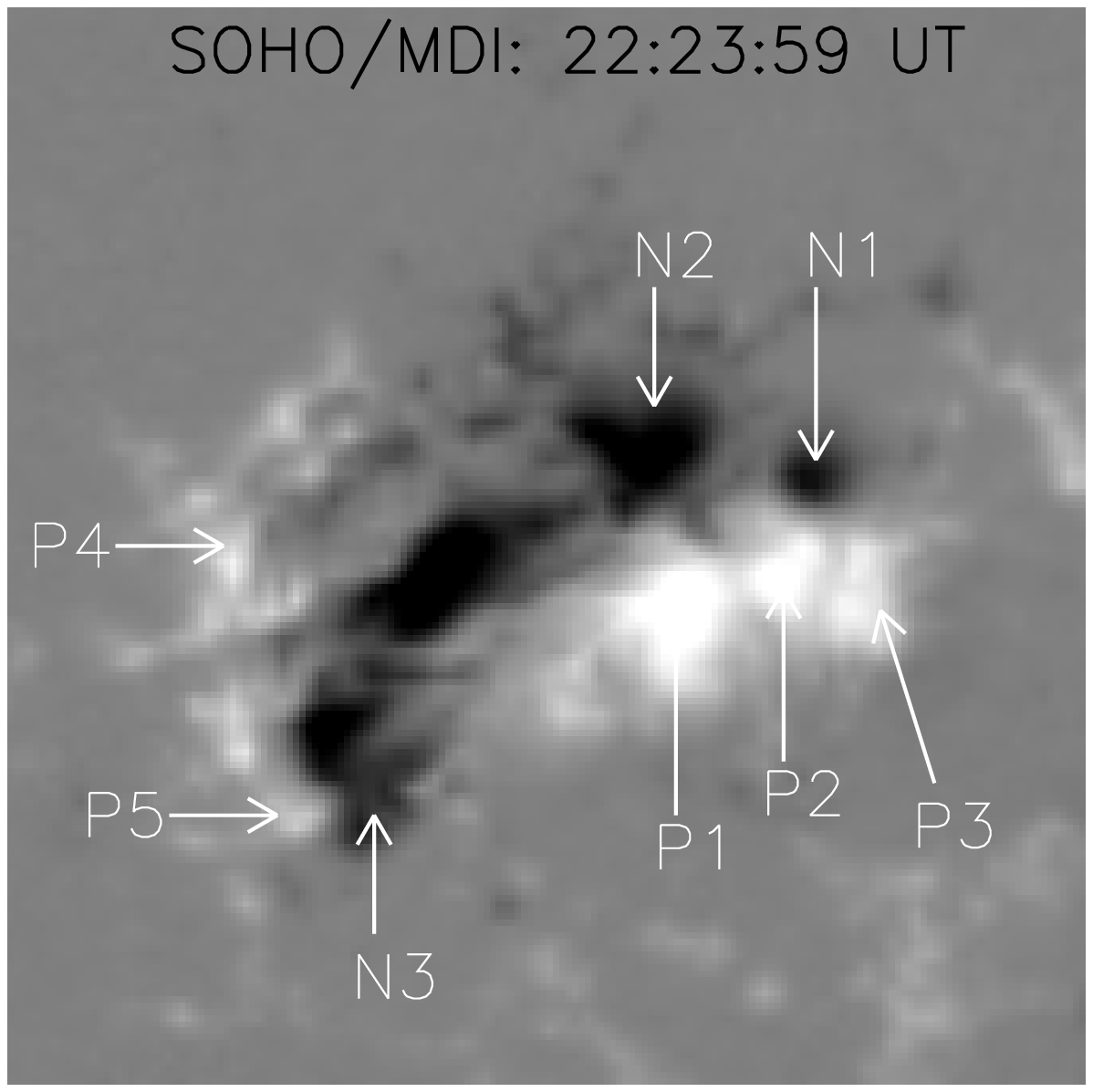}
$\color{black} \put(-160,145){\bf\large (b)}$         
              }
     \centerline{\Large \bf     % Includes the labels (here needs the color
                                %   package, see beginning of this file)
         \hfill}
\vspace{-0.02\textwidth}    % Shift back to the panel bottom

\caption{ SOHO/MDI magnetogram of the active on 27 May, 2003 at two 
selected times. The polarities
associated with the motions described in the text
are shown by arrows ( P1, P2, P3, P4, P5: positive polarity; N1, N2, N3: negative polarity).
The field-of-view of the images is 160$''\times 160''$ . North is up and West is to the right.}

   \label{mdi1}
   \end{figure}

%%%%%%%%%%%%%%%%%%%%%%%%%%%%%%%%%%%%%%%%%%%%%%%%%%%%%%%%%%%%%%%%%%%%%%%%%%%%%%%%%%%%%%%%%%%%%

The emergence of new bipole with high intensity field occurred between
May, 24 and 25 and was well visible inside
the positive polarity area in the magnetograms of May 
25; the bipole had a north-south orientation
(the negative polarity in the north and the
positive polarity in the south). Due to this continuous emergence of
flux in the North part of the active region, the amplitude of the
positive field increased; this positive region progressively surrounded the
new emergent negative flux creating a $\delta$-spot configuration. 
The active region polarities rotated anticlockwise.
We measure the rotation of the magnetic inversion line from May, 25 to May, 27 
and find that of 25 degree.
To draw the magnetic inversion line on the SOHO/MDI magnetograms, we plot 
the simplified contour of zero magnetic field
strength. Afterwards we measure the change in angle of the magnetic inversion line between
25 May, 2003 (06:26 UT) and 27 May, 2003 (06:23 UT).

\begin{figure} %%%%%% FIGURE 5
\centerline{\hspace*{0.008\textwidth}
               \includegraphics[width=0.50\textwidth,clip=]{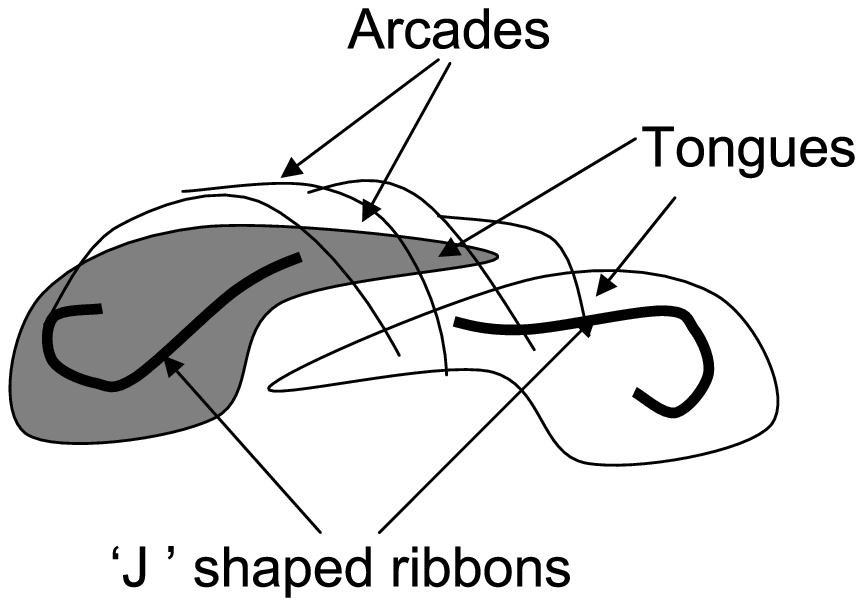}
$\color{black} \put(-100,130){\bf\large (a)}$         
               \hspace*{0.05\textwidth}
\includegraphics[width=0.40\textwidth,clip=]{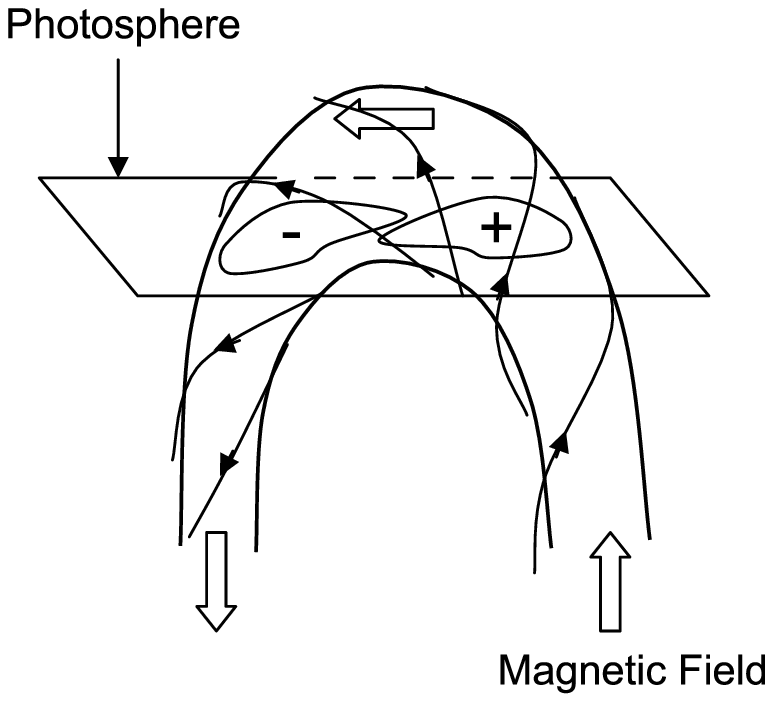}
$\color{black} \put(-80,130){\bf\large (b)}$         
              }
\caption{Sketch of the signatures of positive magnetic helicity
observed in the active region NOAA 10365  during the M1.6 flare on May 27 2003.}
\label{tongue}
\end{figure}

The emergence of the new magnetic flux is a long-term process.
Looking at Figure \ref{mdi}, between May 25 and 27, we see
that the longitudinal magnetic field pattern of the emerging
bipole is  characterized by the existence of two elongated  
areas of opposite polarity labeled N1,2,3 (negative polarities) and P1,2,3 (positive polarities).
These two areas are shifted from one another along the inversion
line (Figure \ref{tongue} (a)). This typical pattern can be explained as 
follows.
The series of magnetograms shows the classical appearance of a
bipole followed by the separation of the two opposite magnetic
polarities as observed for the emergence of an untwisted $\Omega$ loop (Figure \ref{tongue} (b)).
An asymmetry appears in the magnetogram during the
emergence of the apex of the loop because of the contribution of the
azimuthal component of the emerging twisted flux tube to the observed vertical component of the
photospheric field. This azimuthal component produces two elongated polarities so-called
``tongues''. The extension of these tongues is directly proportional
to the magnitude of the twist and their position depends on the sign
(positive twist in our case). The tongues are present only when the
apex of the flux tube is crossing the photosphere during flux
emergence. The positive/negative polarity is elongated towards
the west/east with tongues. This pattern is a characteristic
feature of a long-term twisted emerging flux tube as first described by
\inlinecite{Lopez00} (see their Figure 5).

According to the geometry of the tongues, we conclude that the
emerging tube has  a right-hand twist, and consequently  corresponds to
a positive magnetic helicity {\cite{Lopez00}}.

\subsection{Flow Pattern of the polarities}

We measure the speed of some selected positive (P1-P5) and negative (N1-N3) spots, which
show large displacements (Figure \ref{mdi1}).
For the measurement of the location of the magnetic polarities on magnetograms we manually track the
center of the each polarity.
Based on the magnetograms at 04:50 UT and at 22:23 UT on May 27, we calculate the
velocity and direction of the  selected spots and report them in Table
1. From these measurements we see that the positive polarities P1, P2, P3
and P4 are moving towards the northwest, while polarity P5
is moving towards south. The negative polarity spots, i.e.
N1 and N2 are moving towards the west, while the spot N3 is going
towards south. The velocities of these spots range
from 0.07 to 0.24 km s$^{-1}$.

\begin{table}
\caption{Movements of selected positive (P)/negative (N) polarities}
\label{T-simple}
\begin{tabular}{cclc}     % define the column alignment
                           % l: left, c: center, r: right
  \hline                   % horizontal line

Sunspot~~~ & direction & velocity (km s$^{-1}$)\\
  \hline
P1~~~~~~~~~~& ~~~~~~~~~~~~~~~~northwest~~~~~~~~~~~~~~~~~&~~~~~~~~0.18 \\
P2~~~~~~~~~~& ~~~~~~~~~~~~~~~~northwest~~~~~~~~~~~~~~~~~&~~~~~~~~0.18\\
P3~~~~~~~~~~& ~~~~~~~~~~~~~~~~northwest~~~~~~~~~~~~~~~~~&~~~~~~~~0.07\\
P4~~~~~~~~~~& ~~~~~~~~~~~~~~~~northwest~~~~~~~~~~~~~~~~&~~~~~~~~0.14\\
P5~~~~~~~~~~& ~~~~~~~~~~~~~~~~south~~~~~~~~~~~~~~~~&~~~~~~~~0.24\\
N1~~~~~~~~~~& ~~~~~~~~~~~~~~~~west~~~~~~~~~~~~~~~~~&~~~~~~~~0.12\\
N2~~~~~~~~~~& ~~~~~~~~~~~~~~~~west~~~~~~~~~~~~~~~~~&~~~~~~~~0.10\\
N3~~~~~~~~~~& ~~~~~~~~~~~~~~~~south~~~~~~~~~~~~~~~~&~~~~~~~~0.22\\
\hline
\end{tabular}
\end{table}

These velocities are in agreement with the local correlation tracking (LCT)
analysis done by \inlinecite{Chae04}.
The photospheric motions of these emerging spots lead to a large
diverging and shearing flow pattern of the magnetic field, 
inducing a non-potential configuration
along the magnetic inversion line. The resulting magnetic shear plays an
important role in the onset of the flare.

%%%%%%%%%%%%%%%%%%%%%%%%%%%%%%%%%%%%%%%%%%%%%%%%%%%%%%%%%%%%%%%%%%%%%%%%%%%%%%%%%%%%
\begin{figure}   %%%%%%%%%%%%%%%%%% FIGURE 6

\vspace{0.05\textwidth}    % Shift back to the panel bottom
                                % includes the two top panels
\centerline{\hspace*{0.035\textwidth}
               \includegraphics[width=0.93\textwidth,clip=]{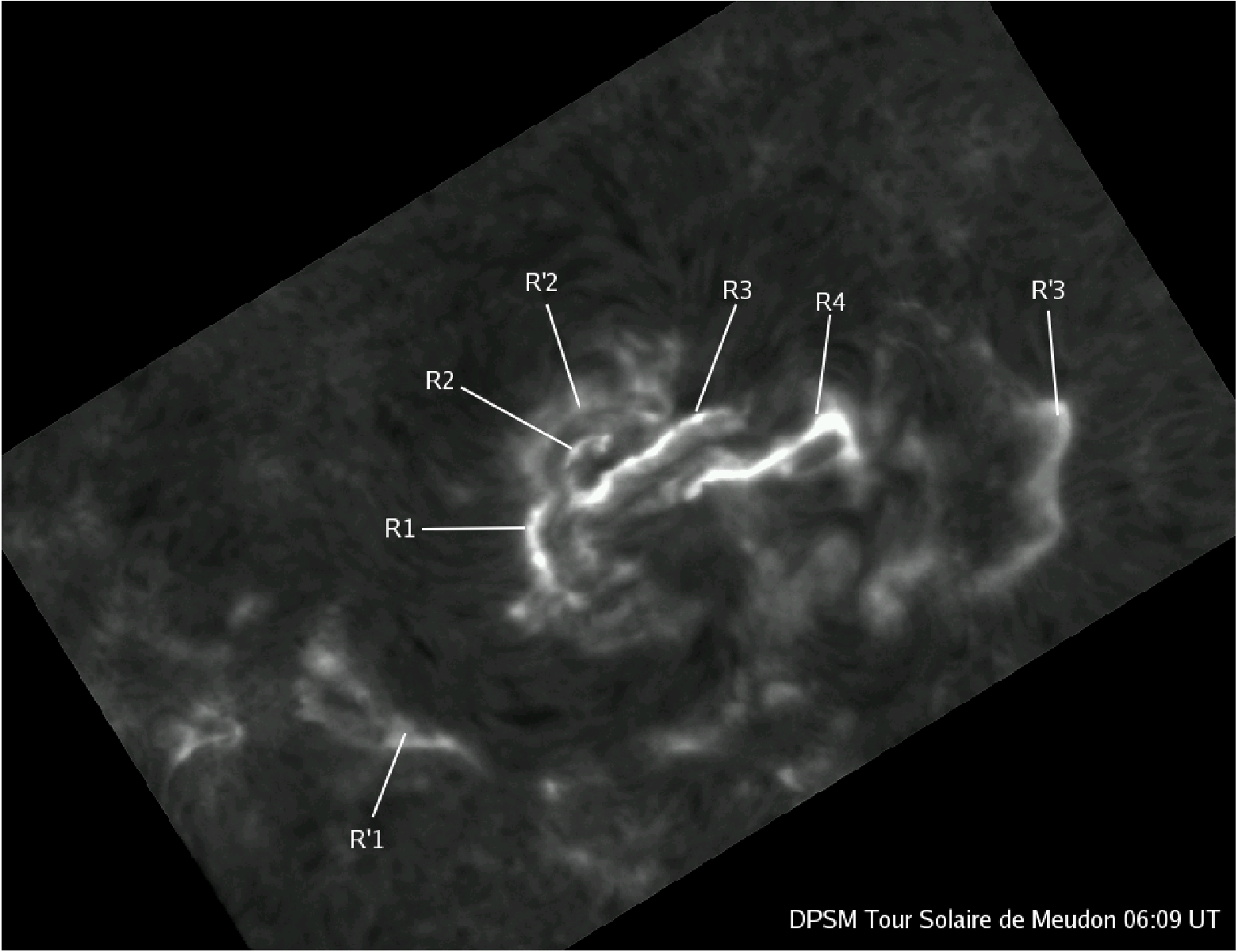}
    \hfill}

\vspace{-0.70\textwidth}   % Shift close to the panel top

     \centerline{\Large \bf     % Includes the labels (here needs the color
\hspace{0.05 \textwidth}  \color{white}{(a)} 
                                %   package, see beginning of this file)
         \hfill}
     \vspace{0.68\textwidth}    % Shift back to the panel bottom
 
  \centerline{\hspace*{0.015\textwidth}
               \includegraphics[width=0.50\textwidth,viewport=50 100 400 348,clip=]{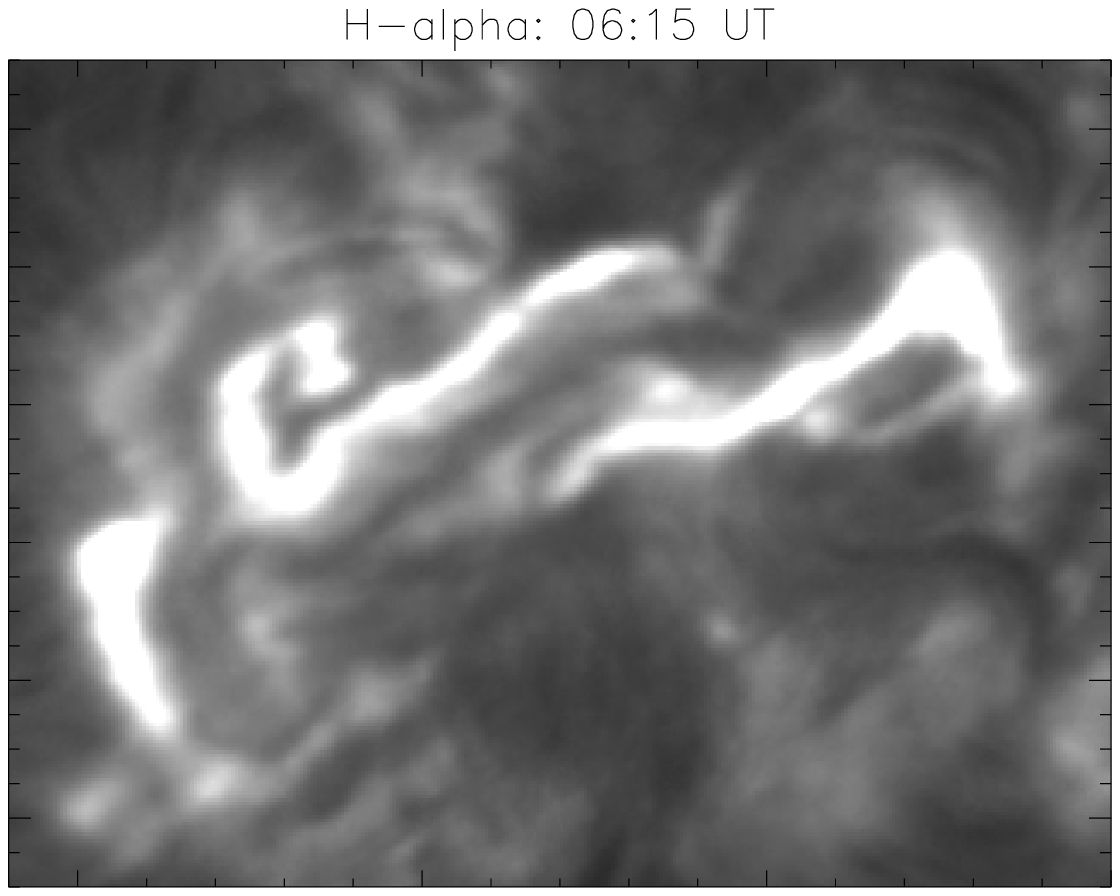}
$\color{white} \put(-160,105){\bf\large (b)}$         
$\color{white} \put(-85,10){\bf H$\alpha$: 06:15 UT}$         
               \hspace*{-0.05\textwidth}
               \includegraphics[width=0.50\textwidth,viewport=50 100 400 348,clip=]{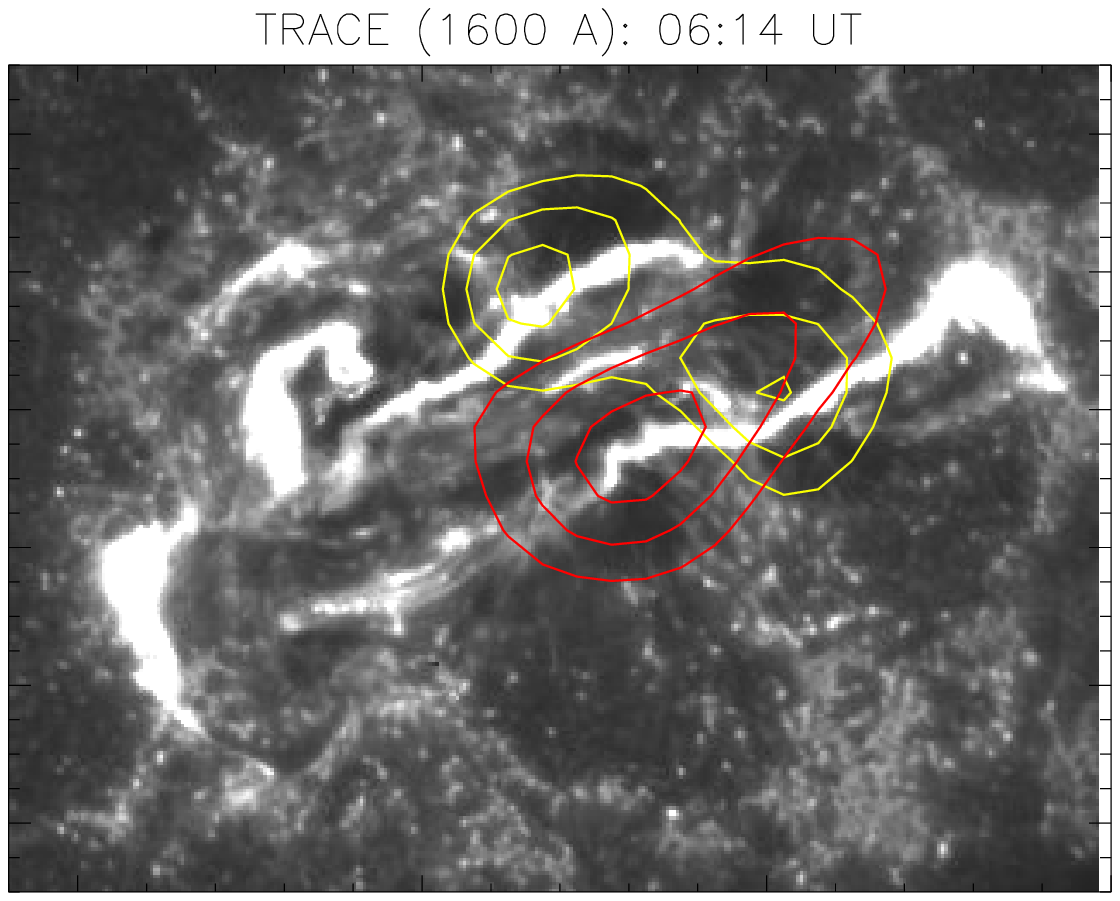}
$\color{white} \put(-160,105){\bf\large (c)}$         
$\color{white} \put(-150,10){\bf TRACE 1600 \AA: 06:14 UT}$         
              }
     \centerline{\Large \bf     % Includes the labels (here needs the color package)
\hfill}
\vspace{-0.05\textwidth}    % Shift back to the panel bottom

\caption{ H$\alpha$ line-center images of the flare observed with the MSDP
spectrograph of Meudon solar tower at 06:09 UT (a, field of view: 456$''\times
290''$ ) and enlarged view of `J' shaped ribbons in H$\alpha$ line-center and TRACE (1600 \AA)
(b, c: field of view: 160$''\times
120''$). The TRACE 1600 \AA\ image is overlaid by RHESSI contours (red contour: 6-12 keV and yellow contour: 25-50 keV, contour levels: 50$\%$, 70$\%$ and 90$\%$ of the peak intensity) at 06:04 UT. The North is up and West is to the right. In this Figure (a), we used the $\it French$ name for the instrument.}
   \label{msdp}
   \end{figure}

%%%%%%%%%%%%%%%%%%%%%%%%%%%%%%%%%%%%%%%%%%%%%%%%%%%%%%%%%%%%%%%%%%%%%%%%%%%%%%%%%%%%%%%%

\section{Description of the Flare} %%%%%%%%%%%%%%%%%%%%%%%%%%%%%%%%%%%%%%%%
      \label{S-general}

The flare was observed in H$\alpha$ and in 1600 \AA\ by TRACE presenting
 `J' shaped ribbons in the central part of the active region and secondary arc-shaped brightenings at the periphery of the active region, as described in 
Section 6:

\subsection{MAIN `J' SHAPED RIBBONS} %%%%%%%%%%%%%%
  \label{S-text}

The flare was observed between 06:09 UT and 06:52 UT in H$\alpha$ with
the MSDP spectrograph in the solar tower of Meudon. We follow 
the evolution of the chromospheric ribbons 
in H$\alpha$. In Figure \ref{msdp} we show the H$\alpha$ reconstructed images in the
line center, indicating the different flare ribbons. The two main flare ribbons are `J' 
shaped. We distinguish four very bright ribbons (R1 to R4) and three secondary 
ribbons, less
bright (R$^\prime 1$ to R$^\prime 3$). R2 and R3 are parts of the same ribbon 
though not formed
exactly at the same time. The pairs of ribbons R1, R2 and R3, R4 are 
part of the main flare; they are located non-symmetrically at both sides of 
the inversion line.
At 06:15 UT, the peak time of the flaring event, 
some brightenings porpagated along the flare ribbons. Such a propagation has
been noticed in previous flares \cite{Berlicki04} and may be explained by a slipping reconnection
mechanism \cite{Aulanier07}.

TRACE observed the central part of the flare in 1600 \AA\ with a
cadence of 6 min. This allows us to identify the central `J' shaped ribbons, which 
were very similar to those  seen in H$\alpha$. An
enlarged view of the two `J' shaped ribbons seen in H$\alpha$ and
TRACE 1600 \AA\ is  presented in Figure
\ref{msdp} (b, c). We shall discuss the `J' shaped ribbons in the
context of emerging twisted flux in Section 5.

\subsection{RHESSI IMAGES AND SPATIAL CORRELATION OF H$\alpha$, TRACE AND HXR SOURCES } %%%%%%%%%%%%%%
  \label{S-labels}

In this section we present  the RHESSI images and the co-alignment
of H$\alpha$, RHESSI, TRACE 1600 \AA\ sources and MDI magnetogram.

\begin{figure}  %%%%%%%%%%%%%%%%%% FIGURE 7
     \vspace{-0.12\textwidth}    % Shift back to the panel bottom
                                % includes the two top panels
   \centerline{\hspace*{0.005\textwidth}
               \includegraphics[width=0.56\textwidth,viewport=0 0 459 368,clip=]{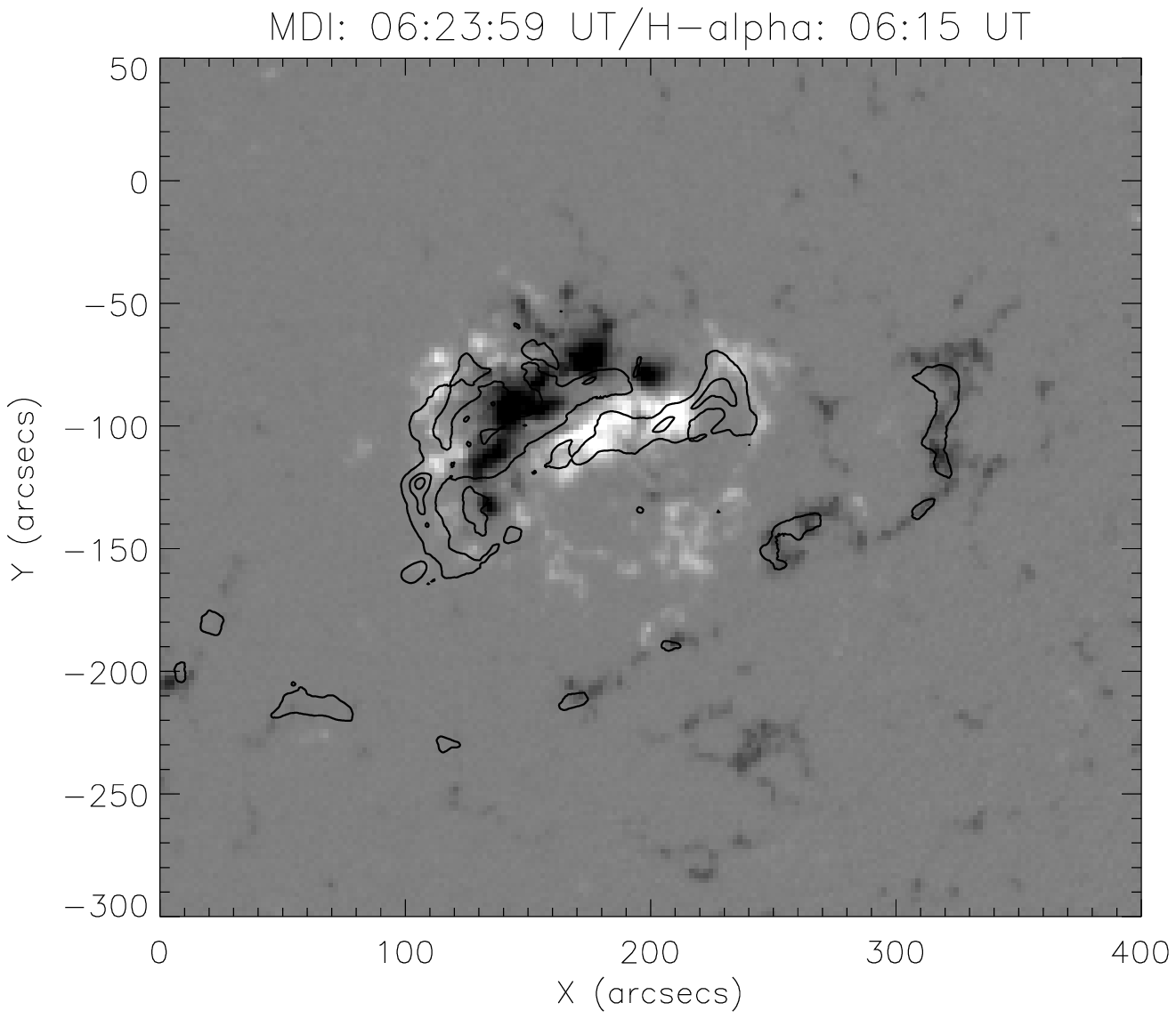}
$\color{white} \put(-165,147){\bf \tiny MDI 06:23/H$\alpha$ 06:15}$         
$\color{white} \put(-170,130){\bf\large (a)}$         
               \hspace*{-0.09\textwidth}
               \includegraphics[width=0.56\textwidth,viewport=0 0 459 368,clip=]{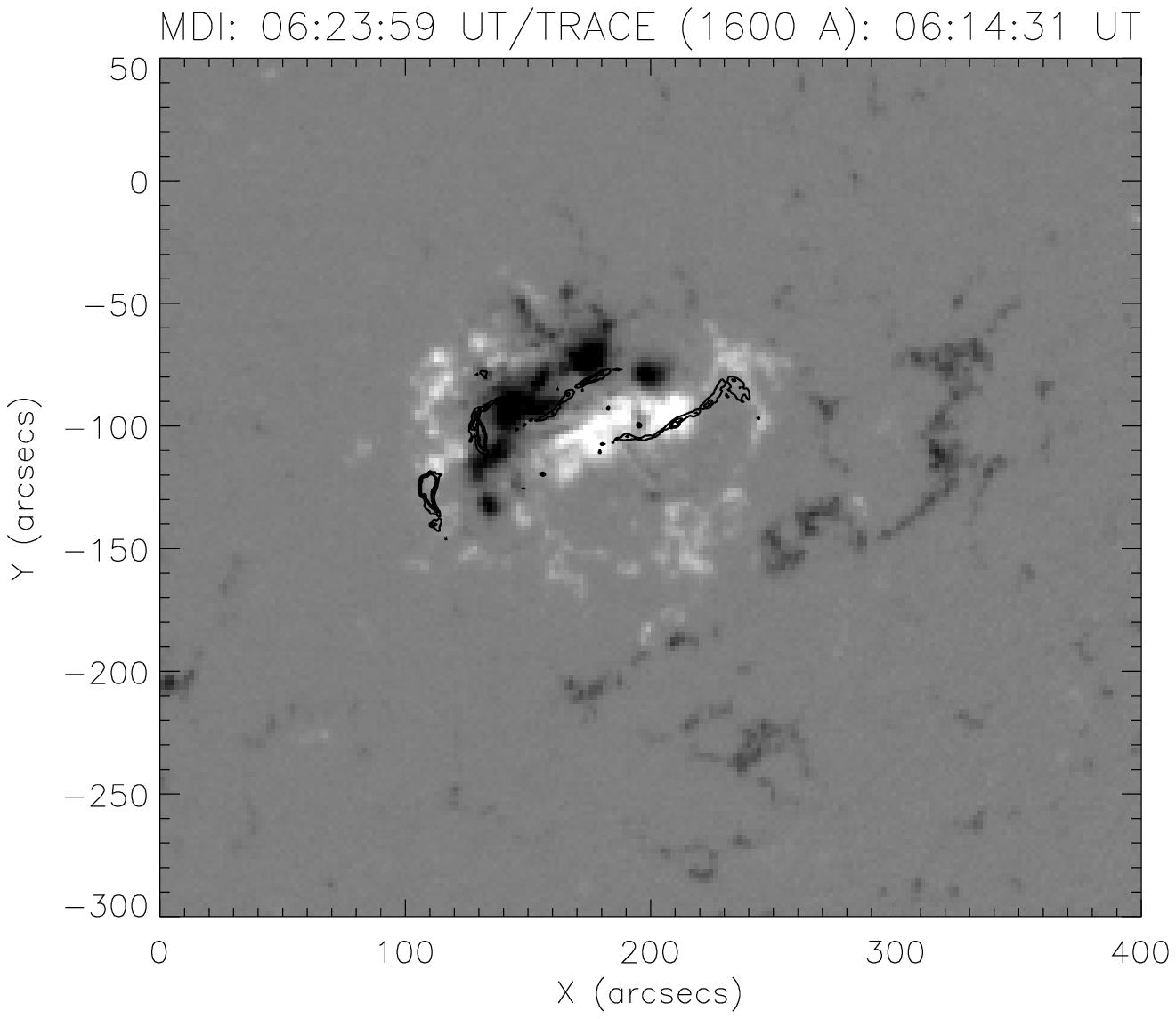}
$\color{white} \put(-166,147){\bf \tiny MDI 06:23/TRACE 1600 \AA\ 01:14}$         
$\color{white} \put(-170,130){\bf\large (b)}$         
              }
     \vspace{-0.05\textwidth}   % Shift close to the panel top
     \centerline{\Large \bf     % Includes the labels (here needs the color
                                %   package, see beginning of this file)
         \hfill}
   \centerline{\hspace*{-0.005\textwidth}
               \includegraphics[width=0.56\textwidth,viewport=0 0 459 368,clip=]{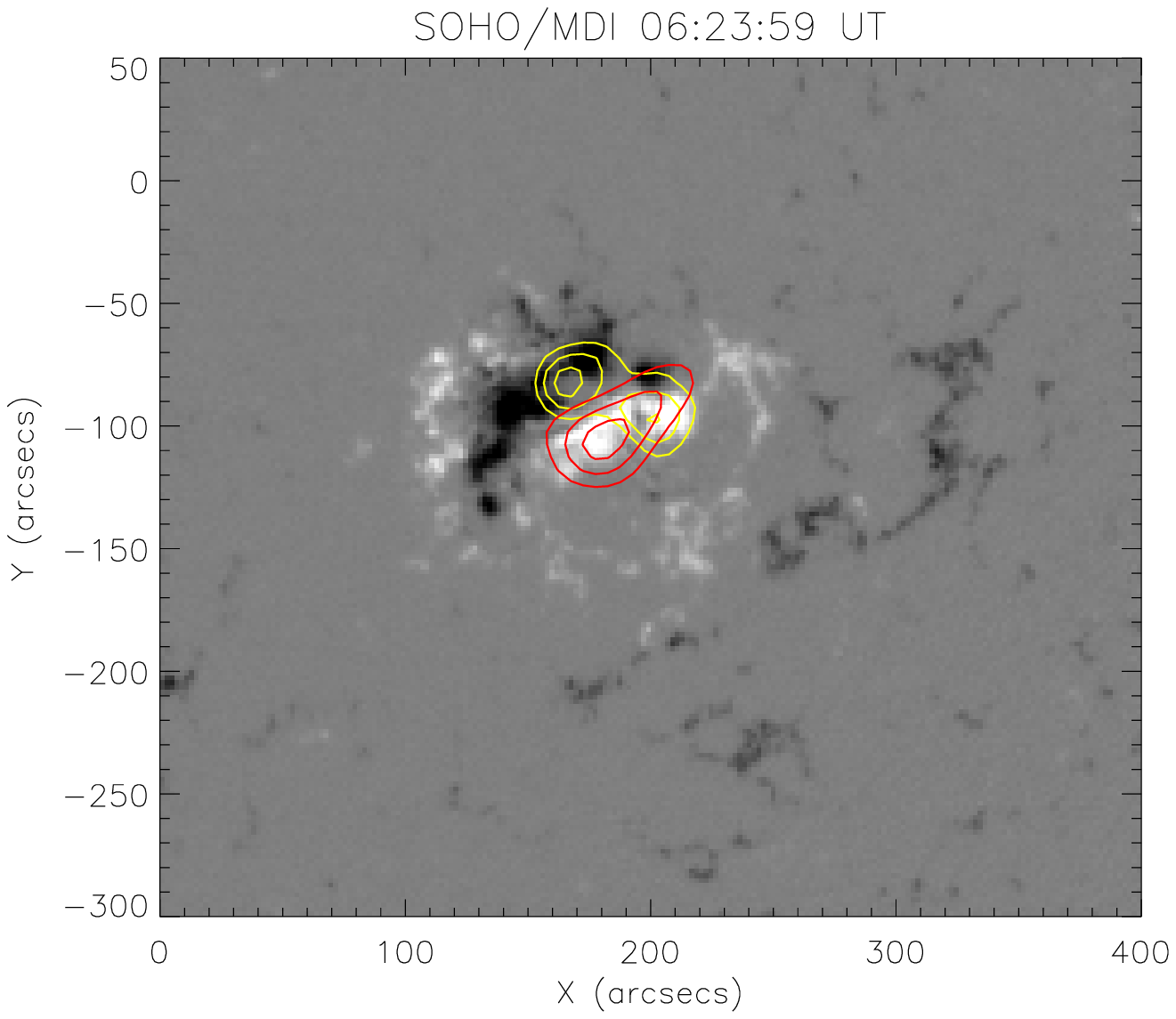}
$\color{white} \put(-165,147){\bf \tiny SOHO/MDI 06:23}$         
$\color{white} \put(-170,130){\bf\large (c)}$         
               \hspace*{-0.09\textwidth}
               \includegraphics[width=0.56\textwidth,viewport=0 0 459 368,clip=]{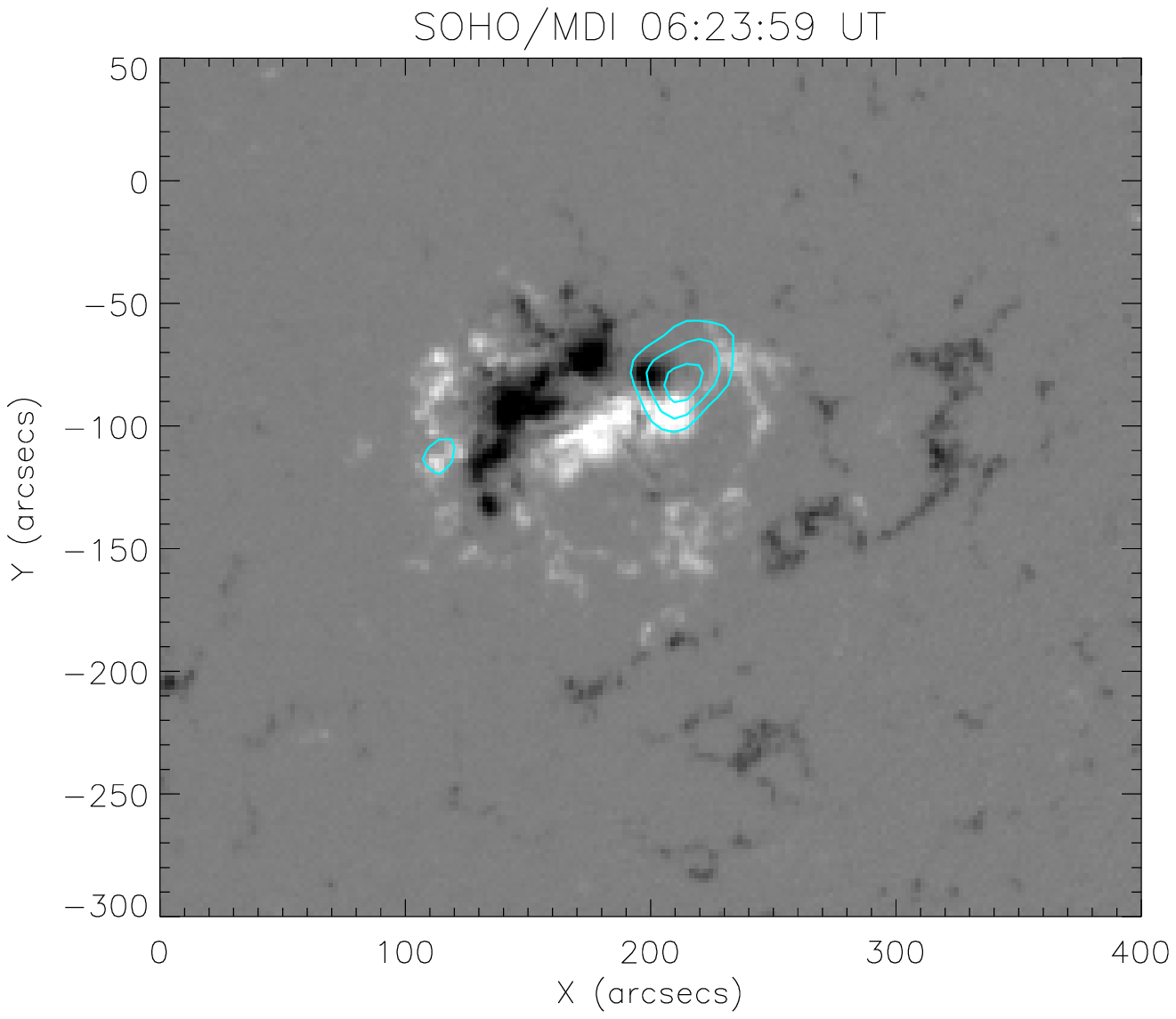}
$\color{white} \put(-165,147){\bf \tiny SOHO/MDI 06:23}$         
$\color{white} \put(-170,130){\bf\large (d)}$         
              }
     \centerline{\Large \bf     % Includes the labels (here needs the color package)
         \hfill}

     %\vspace{-0.2\textwidth}    % Shift back to the panel bottom
     \vspace{-0.05\textwidth}    % Shift back to the panel bottom
   \centerline{\hspace*{0.015\textwidth}
               \includegraphics[width=0.55\textwidth,viewport=50 0 459 308,clip=]{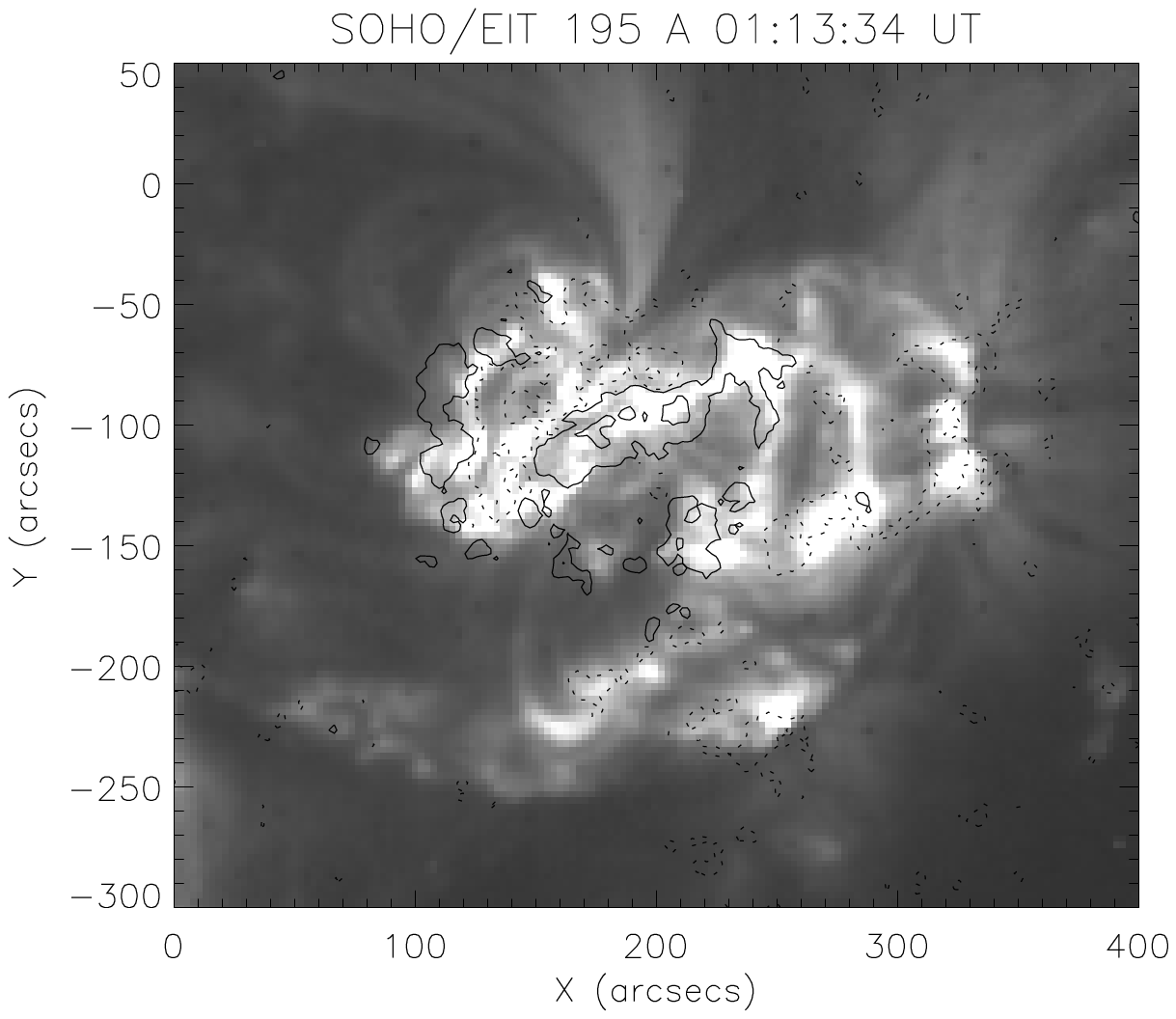}
$\color{white} \put(-160,135){\bf \tiny SOHO/EIT 195 \AA\ 01:13}$         
$\color{white} \put(-145,60){\vector(1,0){25}} \put(-160,57){\L1}$
$\color{white} \put(-170,120){\bf\large (e)}$         
\hspace*{-0.1\textwidth}
               \includegraphics[width=0.55\textwidth,viewport=50 0 459 308,clip=]{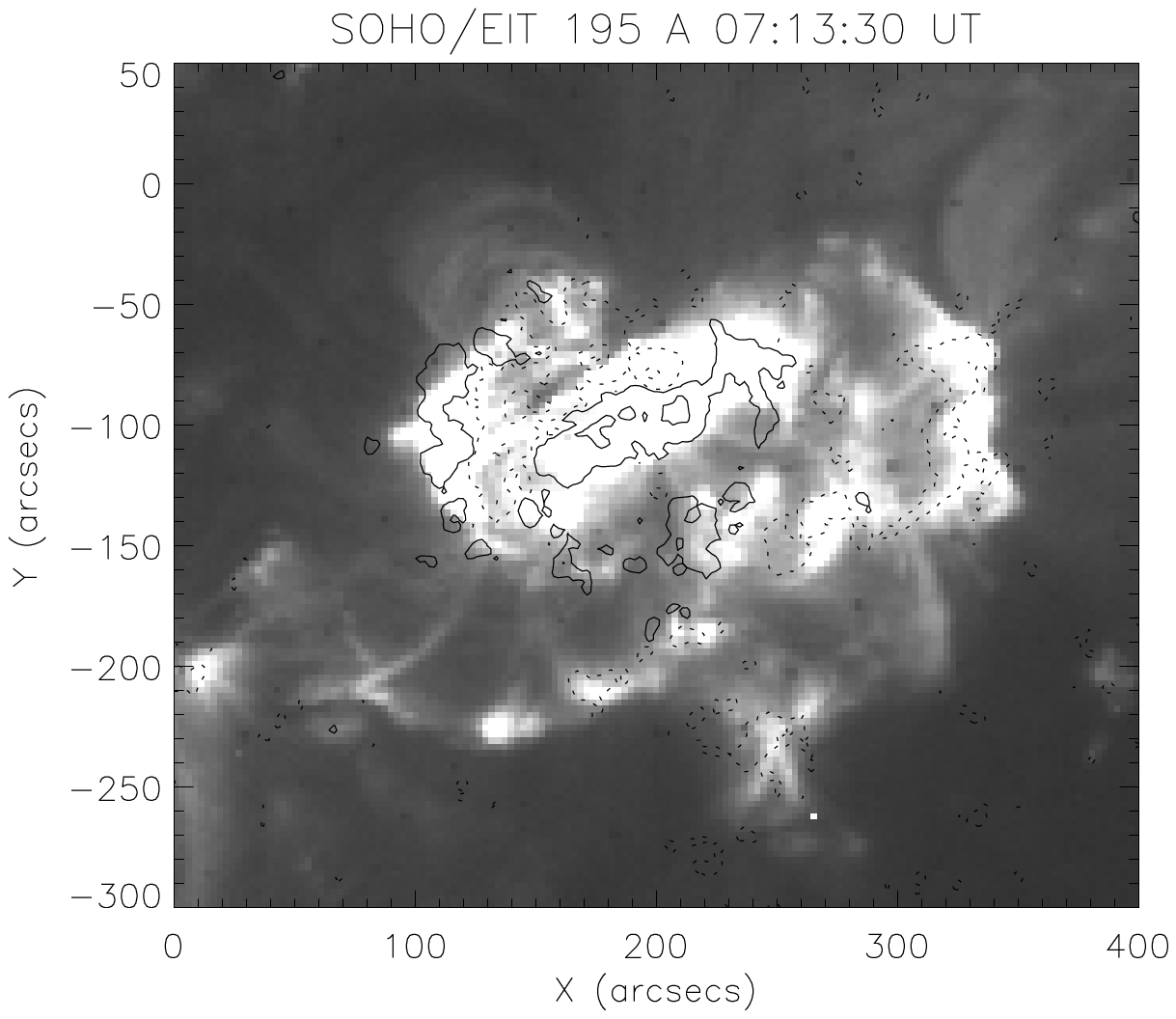}
$\color{white} \put(-160,135){\bf \tiny SOHO/EIT 195 \AA\ 07:13 UT}$
$\color{white} \put(-135,40){\vector(0,1){20}} \put(-140,28){\'L1}$
$\color{white} \put(-170,120){\bf\large (f)}$         
}
     \vspace{-0.08\textwidth}   % Shift close to the panel top
     \centerline{\Large \bf     % Includes the labels (here needs the color package)
         \hfill}
\caption{  Co-alignment of the H$\alpha$ and TRACE 1600 \AA\ ribbons with SOHO/MDI magnetogram
(a, b), MDI magnetogram overlaid by RHESSI contours (6-12 keV (red contour), 25-50 keV
(yellow contour), 15 -20 keV (cyan contour) during the flare impulsive ( 06:04 UT) and 
maximum phase( 06:15 UT), contour
levels: 50$\%$, 70$\%$ and 90$\%$ of the peak intensity) (c, d).
The bottom panel presents the
SOHO/EIT images before (e) and after (f) the flare
onset in 195 \AA . The images are overlaid by MDI magnetogram contours (solid line: positive
polarity, dotted line: negative polarity, contour levels are $\pm 100,\pm1000,\pm1500$ G ). The
loops connecting a secondary ribbon to the main flare ribbon are represented 
by L1 and L$^\prime1$ respectively. North is up and West is to the right.}
   \label{ha_mdi}
   \end{figure}

The co-alignment of RHESSI and MDI data was done by referring the relative position 
of the solar disk
center of each field of view. As for the H$\alpha$ and TRACE (1600 \AA) data, we compare these images with
white-light images of SOHO/MDI.

We reconstructed RHESSI images in 6-12 keV, 15-20 keV and 25-50
keV energy channels from collimators (4F to 9F) using the CLEAN
algorithm \cite{Hurford02}, which gives a spatial resolution of about 12 arcsec.
The locations of RHESSI HXR sources in 6-12 keV and 25-50 keV overlaid on TRACE
1600 \AA\ are shown in Figure \ref{msdp} (b).

At 06:04 UT RHESSI observed two non-thermal X-ray footpoint
sources in the 25-50 keV energy band, located over the main ``J''
shaped ribbons anchored in the positive and negative
polarities (c.f. Figure  \ref{msdp} (b) and Figure
\ref{ha_mdi}). The 6-12 keV source corresponds to a loop-top source, which
looks shifted towards the south. This may be due to projection
effects. During the flare peak time (06:15 UT) RHESSI observed also
two HXR sources 15-20 keV, one was located at the same place as
one of the  previous sources (thus it could be a recurrent phenomenon) and
the other one was located in the east part of the region. The new
source on the left was related to the brightening
of ribbons R1 and R2 at 06:15 UT.

There are several different models for the acceleration of the
non-thermal electrons responsible for HXR emission in the 25-50 keV
range. Reconnection mechanism is commonly proposed.
Other models include the electric field generated in
a reconnection region, in shocks or  turbulence. Following acceleration, electrons
precipitate along the magnetic field to denser layers of the
atmosphere. Electron energy loss in the chromosphere due to Coulomb
collisions leads to heating, causing evaporation along the loops,
while electron-ion bremsstrahlung leads to HXR emission.

Figure \ref{ha_mdi} shows the result of the co-alignment of
H$\alpha$, TRACE 1600 \AA, RHESSI HXR 15-20 keV sources with MDI
magnetograms.
From these figures, we can infer that a strong shear is present in the flare region 
since the ribbon
pairs R1-R2 and R3-R4 are located two-by-two in zones of opposite
polarity in a non symmetric way. The loops associated with these
ribbons are not in a potential state, otherwise their 
footpoints would be located symmetrically at both sides of  
the magnetic inversion line. The co-alignment
permits us to explain the position of the ribbons, their shape and
also how their propagation is stopped where the plages have
different polarities.

\begin{figure}   %%%%%%%%%%%%%%%%%% FIGURE 8

\vspace{-0.80\textwidth}    % Shift back to the panel bottom
                                % includes the two top panels
 \centerline{\hspace*{0.35\textwidth}
               \includegraphics[clip=]{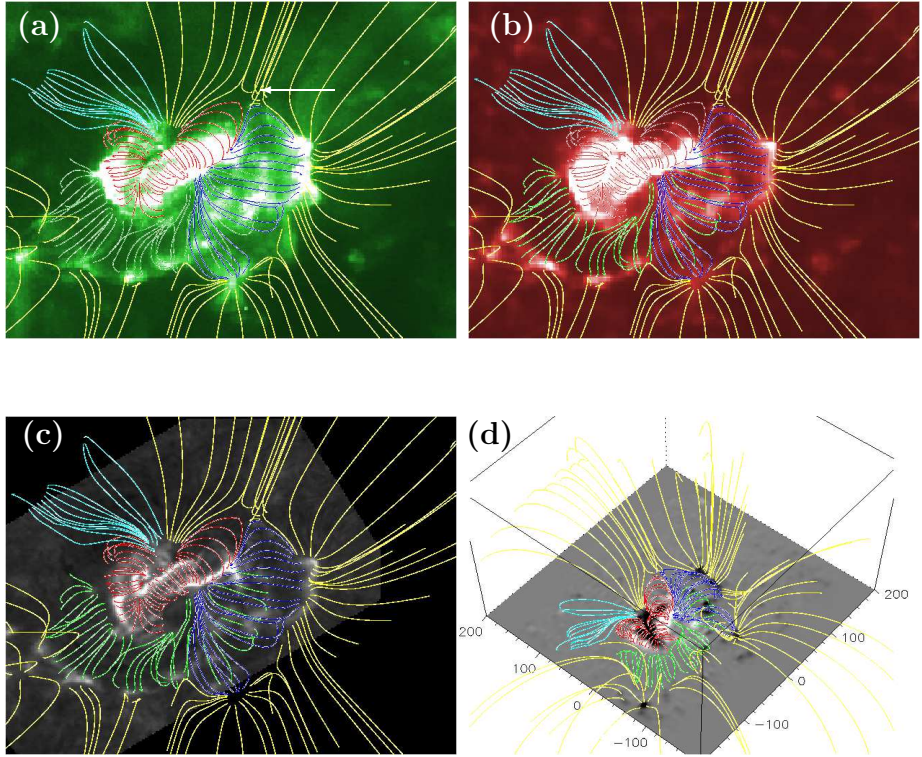}
$\color{white} \put(-530,555){\bf\large (a)}$         
$\color{white} \put(-355,555){\bf\large (b)}$         
$\color{white} \put(-535,400){\bf\large (c)}$         
$\color{black} \put(-370,400){\bf\large (d)}$         
}

\vspace{-0.8\textwidth}    % Shift back to the panel bottom

\caption{SOHO/EIT 195 \AA\ (07:13:30 UT; a), SOHO/EIT 304 \AA\ (07:19:17 UT;
b), H$\alpha$ (06:15 UT; c) images overlaid by magnetic field lines
resulting from a potential field extrapolation.
SOHO/MDI magnetogram at 06:24:11 UT 
and 3D view of magnetic field lines computed from a potential field model.
(d). The location of null point is shown
by an arrow in the upper left panel. }

\label{extra}
\end{figure}

\section{Magnetic Helicity of the Emerging Flux and Flare Ribbons}

The magnetic helicity is a topological quantity that measures the torsion, the
shear, and the twist of magnetic field lines in a given volume \cite{Berger84}. Its definition is: H = $\int \avec \cdot \bb \dv$, where {\bf A} is the
potential vector of the magnetic field {\bf B}.

The magnetic helicity injection has been computed for different regions by
\inlinecite{Chae04} and \inlinecite{Pariat06} using different
methods. Recently \inlinecite{LaBonte07} did a statistical study relating  the occurrence of big flares and found that the big flares need a high magnitude of helicity flux.
\inlinecite{Chae04} present a detailed study of the rate of
helicity ejection in this active region during its passage across
solar disk. The magnetic helicity of our region is found to be positive.
The rate of helicity ejection peaked on 27 May, 2003,
during the main activity of the active region, with the value 2.5
$\times 10^{41}$ Mx$^2$ per hour. This is a strong evidence that the
emergence of magnetic flux is an efficient process for magnetic
helicity injection. The CME observed at 06:50 UT associated with this M1.6
flare very probably ejects part of this injected helicity in the heliosphere.

In Section 3 we have described in detail the emerging region. The
diverging flow pattern of the emerging flux (NW for Bz $<$ 0 and SE
for Bz $<$ 0, Bz being the normal component of the magnetic field),
and the shape of the Bz patterns (especially in the Bz$>$0 region), both
are highly suggestive of magnetic ``tongues'' (Figure \ref{tongue}).
The observed longitudinal magnetic field is approximately Bz because the region 
is close to the
disk center. ``Tongues'' with such an orientation and evolution indicate the 
emergence of a flux
tube with right hand twist, i.e. positive magnetic helicity, H$>$0 \cite{Lopez00}.
This sign corresponds to the hemispheric helicity rule discussed by
\inlinecite{Pevtsov95}, who found  that $\sim$ 70 $\%$  of southern
active regions have H$>$0, as the region studied by us. 

In Section 4, we have shown the flare ribbons observed in H$\alpha$
and  in 1600 \AA. In the central part of the region we distinguish
two main forward ``J'' shaped ribbons (Figure \ref{tongue}). Such a behaviour has been already observed by \inlinecite{Moore95} and interpreted by \inlinecite{Demoulin96}.
Using the concept of QSLs in 3D magnetic configurations, 
\inlinecite{Demoulin96b}; \inlinecite{Demoulin97}
have shown that the intersection of QSLs with the chromosphere determines elongated 
regions that match the shape of flare ribbons. In 
\inlinecite{Demoulin96}, it was shown that QSLs computed in a bipolar configuration have 
a 'J' shape at the chromosphere. The forward `J' /reverse `J' shaped ribbons 
indicate that the flare is triggered by coronal
reconnection below a twisted flux tube of positive/negative helicity
(\opencite{Demoulin96}; \opencite{Williams05}) . In our case of forward `J'
shaped ribbons, the twisted flux tube should have a positive helicity.
Note that this QSL interpretation is also consistent with the propagation of 
brightenings (see Section 4.1),
in the frame of slipping or slip-running reconnection \cite{Aulanier07}.

For the same event we find a good agreement between the magnetic helicity sign 
of the emerging
flux and the  signature of reconnection in the corona (Figure \ref{tongue}).
This kind of association was never found before
in a targeted study of a particular region.
\inlinecite{Williams05} analyzed the eruption of a kink-unstable
filament accompanied by a flare having 'J' shaped ribbons
but did not describe the
emergence phase of the region. Regarding the ``tongues'', only a few papers
have described such phenomena (\opencite{Li07},
\opencite{Green07}), and they have not associated them with flare signatures.

\section{Large Scale Magnetic Field Configuration} %%%%%%%%%%%%%%%%%%%%%%%%%%%%%%%%%%%%%%%%
      \label{S-features}

At the periphery of the active region 10365, secondary
arc-shaped brightenings are observed at the beginning of the main flare on
27 May, 2003. These secondary ribbons are visible  in H$\alpha$ (
R$^\prime 1$,  R$^\prime 3$ in Figure 6) and in EUV (SOHO/EIT Figure
7)

At the periphery of the active region, we observe loops in 195 \AA\ connecting 
these secondary ribbons with the central part of the
region and it is clear that they changed their directions during the flare (see
the direction of loop L1 before the flare and the direction of loop
L$^\prime 1$ after the flare)(Figure \ref{ha_mdi} (e,f). The
secondary ribbons do not belong to the main flare, they correspond
to the loops connecting relatively stable polarities of the
network.

Active region NOAA 10365 is highly sheared close to the main
inversion magnetic line. However, at large-scales in the
active region we infer that the corona is in a near  potential
state, as is frequently observed far from magnetic inversion lines \cite{Schmieder96}.

The H$\alpha$ and EUV secondary ribbons can be
interpreted by means of a topological analysis of the coronal
magnetic field above the active region NOAA 10365 calculated from a
lfff model $\vec{\nabla}\times\vec{B}=\alpha\vec{B}$. For simplicity, we used  a potential field model, with $\alpha$=0.  This approximation is
justified since we are dealing with fields that are far from the
main current-carrying emerging and erupting twisted fields, the
latter can not be modelled in this approximation.
Therefore we do not model the `J'-shaped ribbons.

Magnetic extrapolations in a potential configuration  show the
existence of null points, separatrices, larger quasi-separatrix layers at high altitudes
that could be responsible for the brightening of the secondary
ribbons (\opencite{Priest95}; \opencite{Demoulin96}; \opencite{Demoulin97}). 
The lfff extrapolation method is the one used in 
the database of FRench Online MAGnetic Extrapolations (FROMAGE).
It is based on the decomposition of the observed longitudinal field 
(z=0) in harmonics using the discrete Fast Fourier Transform. 
The method was first described by \inlinecite{Alissandrakis81}.

Figure \ref{extra} presents the results of the extrapolation
overlying the EIT 195 \AA\ image at 07:13:30 UT, the EIT 304 \AA\ image at
06:24:11 UT, and MSDP H$\alpha$ image at 06:15 UT. All are co-aligned
with the MDI magnetogram at 06:34:11 UT. The last panel represents a 3D
view of the extrapolated magnetic field lines in a box  smaller
than the computation box, to see better the connectivities. Note that since the
magnetic field lines drawn in red in the central part  of the region
are not reliable, this area is strongly sheared due to the rotating
emerging flux and the potential approximation is not appropriated, as 
we have mentioned above. 

The magnetic field lines joining places out of the active regions
are drawn in yellow. These field lines are rather arbitrary because of
the  effects of the periodicity of the method and because of the negative flux 
imbalance of the active region.

The green/dark blue  lines correspond to loops connecting the
secondary ribbon R$^\prime1$/R$^\prime3$  to the eruptive area in the center of the
active region. Though not observed, we consider that these large-scale and weakly sheared loops exist because of the continuous presence of the secondary ribbons.
The light blue lines are associated to ribbon R$^\prime2$.

The results of the potential field extrapolation 
show what follows.
The ribbons R$^\prime1$ and R$^\prime3$ visible in the EIT images and in H$\alpha$
(MSDP) are well explained using this approach.
The lines with the ends in these positive
polarities and going towards the periphery of the region  map well
the loops directed toward the West and the South in the images of
EIT/171 \AA (see Figure \ref{ha_mdi} (e, f) and \ref{extra} (a, b)).

The  pattern of field lines linked to R$^\prime2$ indicates that  this
zone is more eruptive than we expected because  numerous lines
connect to the main flare. Some blue lines are also linked to a small bright plage.

The overlying field lines indicate the possibility of the existence of a
null point where some high altitude coronal reconnection could have
occurred, i.e. in the zone between the red, dark blue and the yellow  lines
in the North of the active region (Figure \ref{extra}, arrow). Indeed, the co-alignment
with the H$\alpha$ MSDP image shows that the secondary ribbons are located at
the footpoints of the lines that come from this zone. Different scenarios 
could explain the triggering of this high altitude reconnection and 
its manifestation as secondary remote ribbons. 
In case of a pre-brightening of the secondary ribbons
before the flare, we could infer that we see a signature of
the breakout model \cite{Antiochos99}
as observed in some events (\opencite{Aulanier00};
\opencite{Mandrini06}). In the case of  brightenings occurring at the
same time as the flare, these could be a remanant of an arc-shaped Moreton 
wave or EIT wave surrounding the active region (\opencite{Moreton60}; \opencite{Thompson98}). The brightenings
observed in our case seem to be
only the stationary part of the EIT waves that \inlinecite{Delannee07} explained 
by Joule heating at the sites of QSLs.
\inlinecite{Balasubramaniam05}
described the Sequential Chromospheric Brightenings (SCB) that appear to be 
different from the Moreton/EIT waves.
They explained these SCB as the footpoints of field lines that extend into 
the corona, where they are energized in sequence by magnetic reconnection 
as coronal fields tear from the chromosphere during the eruption.

%%%%%%%%%%%%%%%%%%%%%%%%%%%%%%%%%%%%%%%%%%%%%%%%%%%%%%%%%%%%%%%%%%%%%%%%%%%%%%%

\section{ Conclusion} %%%%%%%%%%%%%%%%%%%%%%%%%%%%%%%%%%%%%%%%
      \label{S-Conclusion}

We studied the M1.6 flare of May 27, 2003, occurring in the
magnetically complex active region NOAA 10365. We used
multi-wavelength observations, which include different instruments:
the  MSDP operating in the Meudon solar tower, RHESSI, TRACE, SOHO/EIT and SOHO/MDI.

The evolution of the photospheric magnetic configuration shows that the
active region NOAA 10365 is characterized by the emergence of a
twisted large magnetic flux tube.
As the twisted flux is rising, diverging flows lead to a peculiar
pattern of the photospheric longitudinal magnetic field. The
elongated polarities, positive and negative, indicate that a right
hand twisted flux tube emerges exhibiting two elongated ``tongues''
\cite{Lopez00}. This implies to an increase of positive helicity, as also noticed
by \inlinecite{Chae04}.

In H$\alpha$ as well as in TRACE 1600 \AA\ we observe
a double `J' shaped ribbon distribution. This
corresponds to a positive magnetic helicity signature.
The asymmetric sheared distribution of the non-thermal HXR sources observed by RHESSI in
the 25-50 keV range also confirms this result.
It is  the first time that we have been able to find
circumstantial evidence for the same sign of
helicity in the flux emergence and the eruption signatures, although 
it has already been predicted that sunspot whorls, tongues,
`J'-shaped ribbons or sigmoids are signatures of magnetic helicity of
the same sign according to their shape (see the review by
\opencite{Demoulin08}).

Regarding secondary arc-shaped ribbons, observed at
the periphery of the active region, we
are able to correlate them with separatrices found in potential
field extrapolation using the code from the database FROMAGE.

Two scenarios are possible to explain the brightenings of the
secondary ribbons: a breakout of the overlying magnetic field
lines before the main flare as \inlinecite{Antiochos99} proposed, and/or some
compression and ohmic dissipation  produced in separatrices during the flare, as
proposed by \inlinecite{Delannee07}
to explain the components of Moreton/EUV waves. Unfortunately, the lack of
high cadence observations with SOHO/EIT and the limited field of view of TRACE do not
allow to conclude if either scenario indeed applies here. Future high cadence 
and full disc EUV images from the upcoming SDO mission should be well suited to 
address this issue.

%%%%%%%%%%%%%%%%%%%%%%%%%%%%%%%%%%%%%%%%%%%%%%%%%%%%%%%%%%%%%%%%%%%%%%%%%%%

%%%%%%%%%%%%%%%%%%%%%%%%%%%%%%%%%%%%%%%%%%%%%%%%%%%%%%%%%%%%%%%%%%%%%%%%%%%
\begin{acks}
 The authors thank  Arnaud B\'erurier who provided all the analysis of this event during
 a four months of training in solar physics, Dr. Pascal D\'emoulin for fruitful discussions and
Dr. S{\"a}m Krucker for his advices in the RHESSI data processing.
We also thank the anonymous referee for his/her constructive comments and 
suggestions, which improved the 
paper significantly.
R.C. thanks the CEFIPRA
for his post doc grant. Financial support by the European Commission through the SOLAIRE Network
(MTRN-CT-2006-035484) is gratefully acknowledged.

\end{acks}

\mbox{}~\\

\end{article}
\end{document}